\documentclass[12pt,a4paper]{article}
 
\pdfoutput=1 

\usepackage{graphicx}
\usepackage{epstopdf}
\usepackage{jcappub}
\usepackage{amsmath,amsthm,latexsym,amssymb,amsfonts,epsfig}
\usepackage{hyperref}
\usepackage{psfrag}
\usepackage[cp1250]{inputenc}
\usepackage[usenames,dvipsnames]{xcolor}

\newcommand{\be}{\begin{equation}}
\newcommand{\ee}{\end{equation}}
\newcommand{\bea}{\begin{eqnarray}}
\newcommand{\eea}{\end{eqnarray}}
\numberwithin{equation}{section}

\title{f(R) and Brans-Dicke Theories and the Swampland}

\abstract{We discuss the {theoretical} viability of $f(R)$ and Brans-Dicke theories of gravity in light of the recent Swampland conjectures. We show that in the case of perfect fluid domination the swampland conjecture conditions may be easily satisfied and therefore we focus on the constraints in the vacuum theory. We derive the conditions for the swampland conjectures in $f(R)$ and Brans-Dicke framework and find that a large portion of the parameter space is not allowed. Furthermore, we analyze particular $f(R)$ models of inflation and dark energy and in their case, we set the allowed range of $R$.}

\author{Michal Artymowski, Ido Ben-Dayan}
\affiliation{Physics Department, Ariel University, Ariel 40700, Israel}
\emailAdd{michal.artymowski@fuw.edu.pl}
\emailAdd{ido.bendayan@gmail.com}

\begin{document} 

\maketitle
\section{Introduction}
$f(R)$ theories \cite{DeFelice:2010aj,Sotiriou:2008rp} are a natural generalization of General Relativity (GR). This is especially true if one views GR simply as a Taylor expansion for small curvature $R$. Counting degrees of freedom, $f(R)$ theory has an additional scalar degree of freedom, usually dubbed, the scalaron. Any $f(R)$ can be mapped at least classically into the Einstein frame GR plus a scalaron with a potential. The best studied exampled of $f(R)$ is the so-called Starobinsky model \cite{Starobinsky:1980te}. Originally, devised to avoid the Big Bang singularity, the Starobinsky model and its variations \cite{Kehagias:2013mya,Ben-Dayan:2014isa,Bamba:2014mua,Artymowski:2015mva,Artymowski:2014gea,Codello:2014sua,Sebastiani:2013eqa,Broy:2016mdc,Broy:2014xwa} turned out to be a good candidate of cosmic inflation \cite{Akrami:2018odb,Array:2015xqh}. 

Recently, it was conjectured that a theory of canonical scalar fields with a potential, written in the Einstein frame, is consistent with Quantum Gravity only if it fulfills at least one of certain inequalities \cite{Obied:2018sgi,Ooguri:2018wrx,Agrawal:2018own,Garg:2018reu,Denef:2018etk,Roupec:2018mbn,Motaharfar:2018zyb,Ben-Dayan:2018mhe,Andriot:2018wzk}:
\begin{eqnarray}
|\nabla V|\gtrsim c \, V, \,  \label{eq:conditionspure} \\ 
\min{\textit{\{Eigenvalues}(V_{ij})\}}<-\tilde{c} \, V \label{eq:conditionspure2}
\end{eqnarray}
where $\nabla V$ is a gradient of the potential in the field space and $V_{ij}$ is the Hessian matrix of partial second derivatives of the potential with respect to the scalar fields. $c, \tilde{c}$ are ${\mathcal O}(1)$ numbers \footnote{In Ref. \cite{Ooguri:2016pdq} the authors argue that $c$ may be arbitrarily small, assuming no critical dS vacua exist{, which makes the swampland conjecture trivial. For further criticism of the swampland conjecture and alternative viewpoints see e.g. \cite{Ben-Dayan:2018mhe,Kehagias:2018uem,Raveri:2018ddi,Akrami:2018ylq,Kinney:2018nny}. Since the publication of the original version of the conjecture \cite{Obied:2018sgi}, there have been several other versions of the swampland conjecture. One may use the methods we apply here to analyze these versions as well.}}. If so, then $f(R)$ theories can also be subjected to this criteria.
In this work we consider the {theoretical} viability of $f(R)$ and Brans-Dicke theories in light of these recent conjectures. First, we find that almost any given matter (except radiation) present in the theory helps to satisfy the inequalities (\ref{eq:conditionspure},\ref{eq:conditionspure2}). Second, we discuss the vacuum case and give a general inequality in section \ref{sec:VphifR}.
The \eqref{eq:conditionspure} inequality is translated into
\be
f(R) \gtrless f(R_\star) \left(\frac{R}{R_\star}\right)^p
\ee
{where $R_\star$ is an initial condition, where the inequality holds for $R > R_\star$} and $p$ depends on the sign of $Rf_R-2f$ and on the value of $c$. Since for negative potential, \eqref{eq:conditionspure} is trivially satisfied, and cannot add any new constraints to $f(R)$ theories, we discuss only the cases where the $f(R)$ theory gives a positive Einstein frame potential. We further discuss some viable models in section \ref{sec:examples}. 
In section \ref{sec:VphiphifR} we discuss the \eqref{eq:conditionspure2} inequality. In sections \ref{sec:BD},\ref{sec:BDVphiphi} we extend the analysis to Brans-Dicke theory and in section \ref{sec:concl} we conclude.  
\\*

In the following  we use the convention $8\pi G = M_{pl}^{-2} = 1$, where $M_{p}\sim 2\times 10^{18}GeV$ is the reduced Planck mass.

\section{General conditions for $V_\phi$} \label{sec:VphifR}

Let us start with the following Jordan frame action of the $f(R)$ theory
\begin{equation}
S = \int d^4x \sqrt{-g} \frac{1}{2}f(R) + S_m(g_{\mu\nu},\chi,\psi,\ldots) \, ,
\end{equation}
where $S_m$ is an action of matter fields (e.g. $\chi$, $\psi$ etc), with total energy density $\rho_m$ and pressure $p_m$. The gravitational part of the action may obtain its canonical GR form in the Einstein frame, where the Einstein frame metric is defined by
\begin{equation}
g^{E}_{\mu\nu} = f_R \, g_{\mu\nu}
\end{equation}
where\footnote{Throughout the paper differentiation will usually be denoted by a subscript.} $f_R\equiv \frac{df}{dR}$,
and the action takes the form of
\begin{equation}
S = \int d^4x \sqrt{-g^{E}}\left(\frac{1}{2}R^{E} + \frac{1}{2}(\partial\phi)^2 - V(\phi) \right) + S_m(g^E_{\mu\nu},\phi,\chi,\psi,\ldots) \, 
\end{equation}
The Einstein frame field $\phi$ and the Einstein frame potential $V(\phi)$ are equal to
\begin{equation}
\phi = \sqrt{\frac{3}{2}} \log f_R \, \qquad V(\phi) = \left.\frac{R f_R - f}{2f_R^2}\right|_{R=R(\phi)} \, . \label{eq:EFpotential}
\end{equation}
 Assuming that the space-time can be described by the flat FLRW metric, the continuity equation for the Einstein frame field gives
\begin{equation}
\ddot{\phi} + 3H^E\dot{\phi} + V_{\phi} = \frac{1}{\sqrt{6}}\left(\rho_m^E - 3p_m^E\right) \, , \qquad \dot{\rho}_m^E + 3H^E\left(\rho_m^E+p_m^E\right) = -\frac{1}{\sqrt{6}}\dot{\phi}\left(\rho_m^E - 3p_m^E\right) \, , \label{eq:contEin}
\end{equation}
where $H^E = \frac{\dot{a}_E}{a_E}$, $a^E = \sqrt{f_R} \, a$, $\rho_m^E = \rho_m/f_R^2$, $p_m^E = p_m/f_R^2$, $a$ is a Jordan frame scale factor and $\dot{\ }$ denotes the derivative with respect to the Einstein frame time defined by $dt^E = \sqrt{f_R}dt$. Note that the RHS of the Eq. (\ref{eq:contEin}) is proportional to the trace of the energy-momentum tensor and therefore the radiation does not generate any source term for the continuity equation of $\phi$ or $\rho_m^E$. From the Eq. (\ref{eq:contEin}) one can see, that $\rho^E$ is not a conserved quantity on the Einstein frame. Therefore, knowing that for $p = w \rho$, where $w = const$, and $\rho = \rho_0 (a_0/a)^{3(1+w)}$ let us introduce 
\begin{equation}
\rho^\star = \rho_0 \left(\frac{a_0}{a^E}\right)^{3(1+w)} \, , \label{eq:rhostar}
\end{equation}
which is a conserved quantity in the Einstein frame. {Note that} even though $\rho^\star$ is an Einstein frame quantity, it scales like a Jordan frame energy density and it is fully $\phi$-independent. It is therefore useful to use it to extract the $\phi$ dependence in $\rho_m^E$. Substituting Eq. (\ref{eq:rhostar}) into (\ref{eq:contEin}) one finds the effective scalar potential of the form of
\begin{equation}
V^{eff}(\phi) = V(\phi) + e^{-\frac{1}{\sqrt{6}}(1-3w)\phi} \rho^\star  = V(\phi) + \rho^E_m\, . \label{eq:Veff}
\end{equation}
{We want to argue that in the non-vacuum case one should indeed take into account the interactions between the scalaron and different fluids, which fill the Universe. As argued in \cite{Obied:2018sgi} the swampland conjecture is phrased in the Einstein frame, therefore in our case the coupling between the scalaron and the matter is relevant when assessing the theoretical viability of the theory.} %in which the scalaron is coupled to the matter and therefore the matter is important for the validity of the conjecture.} %One can still argue that the swampland conjecture was designed to constrain theories coupled only to gravity. Nevertheless, in the case of the scalaron any interaction with other fields is in fact a gravitational one, since the scalaron acts as a field, which mediates the gravitational interaction.} 
Within this framework from Eq. \eqref{eq:conditionspure} it implies that 
\begin{equation}
|V^{eff}_\phi| > c \,  V^{eff}\, , \label{eq:conditions}
\end{equation}
where $V^{eff}_{\phi} = \frac{dV^{eff}}{d\phi}$ and $c \sim \mathcal{O}(1)$. The $V^{eff}>0$ condition is equivalent to 
\begin{equation}
Rf_R + 2\rho_m > f \, . \label{eq:Vpositive}
\end{equation}
In addition we require that
\begin{equation}
f_R > 0 \, , \qquad f_{RR} > 0 \label{eq:ghost}
\end{equation}
where $f_R>0$ guarantees that the gravitons are not ghostly and that the transition to the Einstein frame is possible and $f_{RR}>0$ secures the positive mass of the curvature fluctuations \cite{DeFelice:2010aj}. Another 'swampland' property is the restriction of the theory to the certain range of $\phi$, otherwise one expects a tower of light states to appear as $\Delta \phi$ grows beyond $M_{pl}$, \cite{Ben-Dayan:2018mhe,Grimm:2018ohb}, hence:
\begin{equation}
\Delta \phi \lesssim 1 \, .\label{eq:deltaphi}
\end{equation}

One can express the derivative of the Einstein frame potential in terms of $R$, $f(R)$ and $f_R$ by noticing, that
\begin{equation}
\frac{dV}{d\phi} = \frac{dV}{dR}\frac{dR}{df_R}\frac{df_R}{d\phi} \, .
\end{equation}
From
\begin{equation}
V_R = \frac{f_{RR}}{2f_R^3}(2f-Rf_R)
\end{equation}
one finds
\begin{equation}
V_\phi = \frac{1}{\sqrt{6}f_R^2}(2f-Rf_R) \, . \label{eq:dVdphi}
\end{equation}
Finally, from Eqs. (\ref{eq:EFpotential},\ref{eq:conditions},\ref{eq:dVdphi}) one obtains 
\begin{equation}
\left|2f-Rf_R - (1-3\omega)\rho_m\right| > \sqrt{\frac{3}{2}} \, c \, (Rf_R-f + 2\rho_m) \, . \label{eq:conditionfR}
\end{equation}

The presence of matter fields may relax the condition on the Einstein frame potential $V(\phi)$. Note, that for the domination of matter fields, i.e. for $V_\phi \ll (1-3w)\rho^E_m$ and $V \ll \rho^E_m$ one finds
\begin{equation}
\frac{|V^{eff}_\phi|}{V^{eff}} \simeq \frac{1}{\sqrt{6}}|1-3\omega| \, , \label{eq:Rhodomination}
\end{equation}
which in the case of Jordan frame cosmological constant or stiff matter is of the order of one. The simplest example, which satisfies the swampland conjecture is the $w \simeq -1$ case of a perfect fluid, which mimics Jordan frame cosmological constant. It is realistic to assume that this perfect fluid is a scalar field. Such a field may be used as an inflaton or as a source of dark energy \footnote{{At this point, we consider any value of $c$, which is of the order of one. Nevertheless, the exact value of $c$ may be crucial for the phenomenological consistency of the model. For details see Refs. \cite{Raveri:2018ddi,Akrami:2018ylq}.}}. One usually assumes that its potential is (at least) locally flat and therefore without the presence of the $f(R)$ it would not satisfy (\ref{eq:conditions}). The $f(R)$ scalaron plays here a role of an additional field with steep potential \cite{Achucarro:2018vey}, which enables flat inflationary potential together with a consistency with (\ref{eq:conditions}). Note that the first derivative of the potential of fields that contribute to $\rho_m$ does not need to be included in our analysis since $V_\phi$ already provides enough of steepness.

We want to emphasize that radiation does not improve the {theoretical} viability of the theory from the point of view of the swampland conjecture. Not only it does not increase the value of $V^{eff}_\phi$, but also it increases the value of $V$, which finally decreases the $|V^{eff}_\phi|/V^{eff}$ ratio.

Since the domination of $\rho_m$ helps to satisfy the swampland conjecture we want to focus on the opposite case, which is $\rho_m \rightarrow 0$. From now on we will assume that $\rho_m$ is negligible. 
Hence to fulfill the first possibility of the swampland conjecture \eqref{eq:conditionspure}, one has to fulfill:
\be
\left|2f-Rf_R\right| > \sqrt{\frac{3}{2}} \, c \, (Rf_R-f) \, . \label{eq:conditionfRvac}
\ee

\subsection{The $Rf_R<2f$ case}

First of all let us assume
\begin{equation}
f<Rf_R<2f \qquad \Rightarrow \qquad 2f-Rf_R >  \sqrt{\frac{3}{2}} \, c \, (Rf_R - f) \, ,
\end{equation}
We assume $f<Rf_R$ because otherwise according to \eqref{eq:EFpotential}, $V^E(\phi)<0$ and the criteria are trivial. This leads to
\begin{equation}
Rf_R < \alpha f =  \frac{2+\sqrt{\frac{3}{2}}\, c}{1+\sqrt{\frac{3}{2}}\, c} f \, . \label{eq:Vphipositive}
\end{equation}
We have assumed that $c>0$ and since then
\begin{equation}
\alpha\equiv\frac{2+\sqrt{\frac{3}{2}}\, c}{1+\sqrt{\frac{3}{2}}\, c} \, \in \, (1,2) \, ,
\end{equation}
where the lower and upper limits correspond to $c \to \infty$ and $c \to 0$ respectively. {The presented range of $c$ covers all possible $c \in (0,\infty)$. Since $c$ need to be of order of unity in order to be consistent with the swampland conjecture, one expects $\alpha$ to  be in vicinity of $\sim 1.4$.}
Therefore the models, for which 
\begin{equation}
f < Rf_R < \alpha f \label{eq:ineq1}
\end{equation}
satisfy initial assumption from the Eq. (\ref{eq:Vphipositive}) and the condition (\ref{eq:conditionfRvac}). On the other hand for
\begin{equation}
\alpha f < Rf_R < 2 f 
\end{equation}
one satisfies only {$R < R f_R < 2f$}, but not the condition (\ref{eq:conditionfRvac}), which makes it an excluded region. Note that $Rf_R<2f$ indicates that $f$ must be positive, since  we have assumed $f_R>0$. 

\begin{figure}[h!]
\centering
\includegraphics[height=7cm]{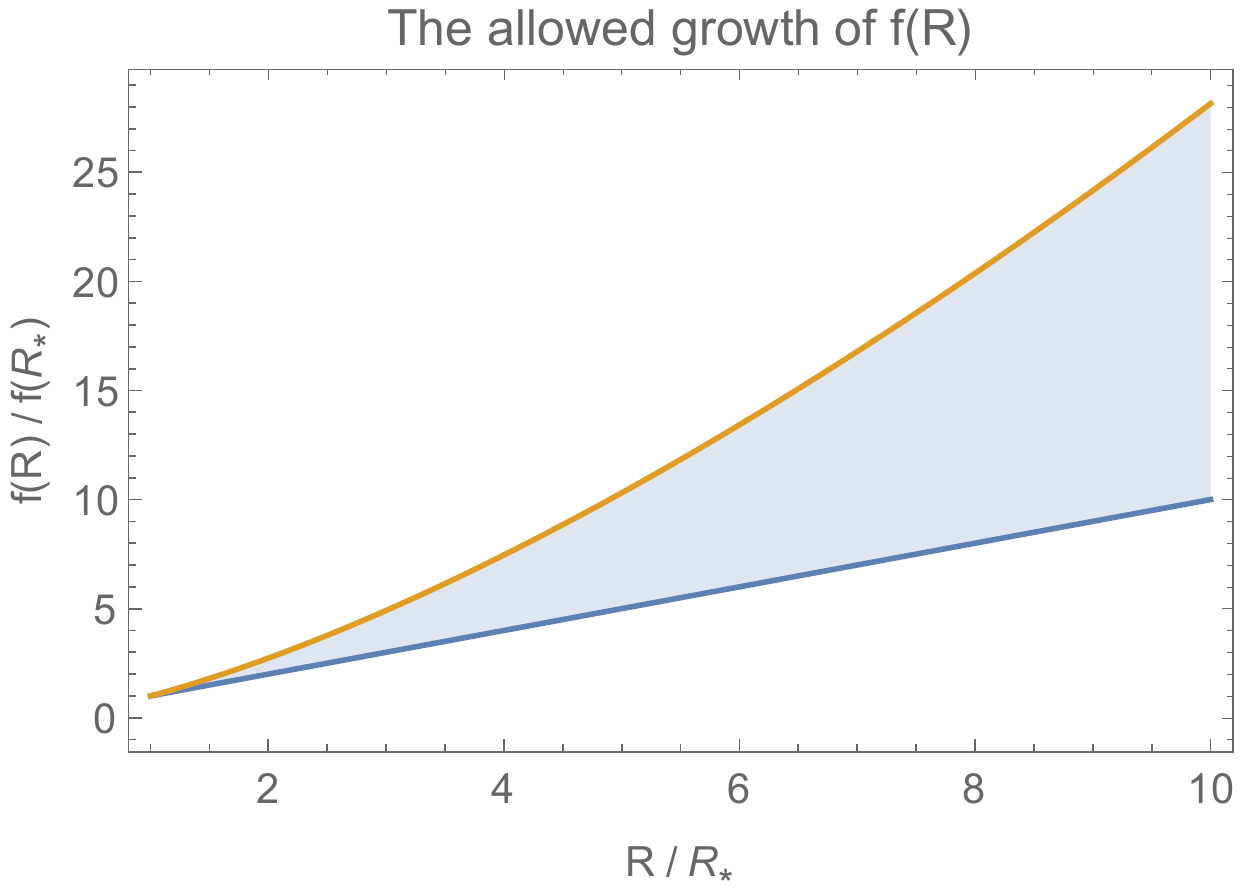}  
\caption{An example of the allowed growth of $f(R)$ for{ $R < R f_R < 2f $ and} $c=1$, as given by equation \eqref{eq:two-sided}. {Lower and upper bounds correspond to $\alpha = 1$ and $\alpha \simeq 1.45$ respectively.} The shaded region represents the allowed space of theories.}
\label{fig:general}
\end{figure}

The general solution for any inequality of the form of
\begin{equation}
f'(R) \lessgtr \theta(R) f(R) \label{eq:generalinequality}
\end{equation}
is
\begin{equation}
f(R) \lessgtr f(R_\star)\exp \left( \int_{R_\star}^R \theta(x)dx \right) \, , \label{eq:ineqsol}
\end{equation}
where $R_\star$ is some initial condition, above which the inequality is valid. As we will see,  the general conditions for $f(R)$ are  $\theta(R)=\frac{\alpha,\gamma}{R}$, where $\alpha$ is defined above and $\gamma$ is defined in  \eqref{eq:caseafR} below. The use of $\alpha$ or $\gamma$ depends on the sign of $Rf_R-2f$ and on the value of $c$. Integrating the inequality \eqref{eq:generalinequality} gives:
\begin{equation}
f(R_\star)\exp \left( \int_{R_\star}^R \theta(x)dx \right) = f(R_\star)\left( \frac{R}{R_\star} \right)^{n} \, , \label{eq:generalineqsol}
\end{equation}
{where $n = \alpha,\gamma$. From the Eq. \eqref{eq:ineq1} one finds}
\begin{equation}
\label{eq:two-sided}
\frac{R}{R_\star}<\frac{f(R)}{f(R_\star)}<\left( \frac{R}{R_\star} \right)^{\alpha} \, .
\end{equation}
 Hence, the swampland conjecture places a severe constraint on $f<Rf_R<2f$ models. It states that $f(R)$ must grow slower than $R^{\alpha(c)}$ ( for $c=1$ it has to grow slower than $R^{1.45}$). An example of the limited allowed growth of $f(R)$ is plotted for the $c=1$ case in Figure \ref{fig:general}.
 
 {An interesting generalization of the \eqref{eq:conditions} is shown in Ref. \cite{Andriot:2018mav}, where the authors conclude, that the swampland conjecture is satisfied for $(2\epsilon)^{q/2}-a\eta > 1-a$, where $q>2$, $a\in (0,1)$ and $\epsilon$ and $\eta$ are slow-roll parameters. In the single field case the inequality takes the form 
\begin{equation}
\left(\frac{V_\phi}{V} \right)^q - a \frac{V_{\phi\phi}}{V} > 1 - a \, .
\end{equation}
For $f(R) \propto R^p$ this inequality is equivalent to {
\begin{equation}
\frac{2(p-2)^2}{3(p-1)^2}\left(\left(\frac{2(p-2)^2}{3(p-1)^2}\right)^{\frac{q}{2}-1} - a \right) > 1-a \, .
\end{equation}
Note that one finds $\frac{2(p-2)^2}{3(p-1)^2}>1$ for $p < \sqrt{6}-1$. In such a case the inequality is satisfied for any values of $a$ and $q$. Keep in mind that $p<1$ would lead to classical or quantum instability of the model. Thus,} the solution of this inequality is $1<p <\sqrt{6}-1$. Quite surprisingly {(and completely accidentally), for $f(R) \propto R^p$} this condition is fully equivalent to the \eqref{eq:two-sided} for $c=1$.}

\subsection{The $Rf_R>2f$ case}

For $Rf_R>2f$ one finds $2f-Rf_R<0$, which from the Eq. (\ref{eq:conditionfRvac}) gives
\begin{equation}
Rf_R - 2f >  \sqrt{\frac{3}{2}} \, c \, (Rf_R-f) \qquad \Rightarrow \qquad Rf_R \left(1-\sqrt{\frac{3}{2}}\, c\right) > f \left(2-\sqrt{\frac{3}{2}}\, c\right) \, . \label{eq:Vphinegative}
\end{equation}
Let us investigate possible solutions of this inequality in three regimes

\begin{itemize}
\item[{\bf a)}] $\sqrt{\frac{3}{2}}\, c < 1$ - in such a case, from the Eq. (\ref{eq:Vphinegative}) one satisfies the (\ref{eq:conditionfRvac}) for 
\begin{equation}
Rf_R > \gamma f \equiv \frac{2-\sqrt{\frac{3}{2}}\, c}{1-\sqrt{\frac{3}{2}}\, c} f \, , \label{eq:caseafR}
\end{equation}
where 
\begin{equation}
\gamma \in (2,\infty) \, .
\end{equation}
 The lower (upper) bound corresponds to $c \to 0$ ($c \to \sqrt{2/3}$). {The presented range of $\gamma$ covers $c \in (0,\sqrt{2/3})$. One of the crucial features of the swampland conjecture is the fact that $c \sim 1$ and since the $1-\sqrt{3/2} \, c$ term is in the denominator, the final value of $\gamma$ strongly depends on the exact value of $c$. For instance for $c \simeq 0.5$ one finds $\gamma \gtrsim 3.6$}. In such a case all $f(R)$ theories (for both, positive and negative $f(R)$), which satisfy (\ref{eq:caseafR}) automatically satisfy the assumption $Rf_R>2f$ together with a condition (\ref{eq:conditionfRvac}). From Eqs. (\ref{eq:ineqsol}) and (\ref{eq:generalineqsol}) one can see, that the solution of \eqref{eq:caseafR} is
\begin{equation}
f(R)>f(R_\star)\left( \frac{R}{R_\star} \right)^{\gamma} \, .
\end{equation}
Theories with $f>0$, which satisfy
$2f<Rf_R<\gamma f$ do not satisfy (\ref{eq:conditionfRvac}) because they do not satisfy \eqref{eq:caseafR} and therefore are excluded. 

\item [{\bf b)}] $1 < \sqrt{\frac{3}{2}}\, c < 2$ - in such a case one finds
\begin{equation}
Rf_R < \gamma f \, , \label{eq:casebfR}
\end{equation}
where 
\begin{equation}
\gamma \in (-\infty,0) \, . 
\end{equation}
The lower (upper) bound corresponds to $\sqrt{3/2}\, c \to 1$ ($\sqrt{3/2}\, c \to 2$). For positive $f$ there are no theories which would satisfy $Rf_R<\gamma f$ and $Rf_R>2f$. However, all of the theories with $f<0$ and $\gamma < -2$ (which corresponds to $\sqrt{3/2}\, c<4/3$) satisfy $Rf_R>2f$ and the condition (\ref{eq:conditionfRvac}), which from Eqs. (\ref{eq:ineqsol},\ref{eq:generalineqsol}) and \eqref{eq:casebfR} gives
\begin{equation}
f(R)<f(R_\star)\left( \frac{R}{R_\star} \right)^{\gamma} \, .
\end{equation}
Case (b) is most likely impossible to realize for any realistic $f(R)$ model. One can see that by noting, that for the polynomial form of $f(R)$, which contains terms like $a_n R^{-n}$, one can satisfy only $f_R>0$ or $f_{RR}>0$, but not both of these conditions at the same time.

\item[{\bf c)}] $\sqrt{\frac{3}{2}}\, c > 2$ - then the inequality (\ref{eq:casebfR}) still holds, but
\begin{equation}
\gamma \in (0,1) \, .
\end{equation}
In such a case none of the theories can satisfy (\ref{eq:casebfR}) and $Rf_R>2f$ simultaneously. 
\end{itemize}

For most of the scalar theories, the particular value of the $c$ constant does not have crucial implications for the allowed region in the parameter space of the theory. {For instance for given $c$ and $V \propto e^{\pm\lambda \, \phi}$ one may always obtain a consistency with a swampland conjecture by assuming $\lambda \geq c$. This is not the case in} $f(R)$ theory with a region of $V(\phi)$, for which $Rf_R > 2f$ (i.e. for $V_\phi < 0$). {Then} the value of $c$ strongly determines, if there is any $f(R)$ theory, which is consistent with the swampland conjecture{, regardless of the parameters of the theory}. We want to emphasize that this is a unique feature of $f(R)$ theories.

\section{Analysis of particular $f(R)$ models} \label{sec:examples}

 The polynomial of $R$ is not just a bound for $f(R)$, it is also one of the simplest and most intuitive forms of $f(R)$, that may satisfy the inequality (\ref{eq:conditionfRvac}). In this section, we present a detailed analysis of this family of $f(R)$ models. On a model to model basis, we will also discuss the allowed field range of the scalaron $\Delta \phi \lesssim 1$ that corresponds to a limit on the allowed Ricci scalar curvature.

\subsection{Examples of $Rf_R<2f$}

\begin{figure}[h!]
\centering
\includegraphics[height=4cm]{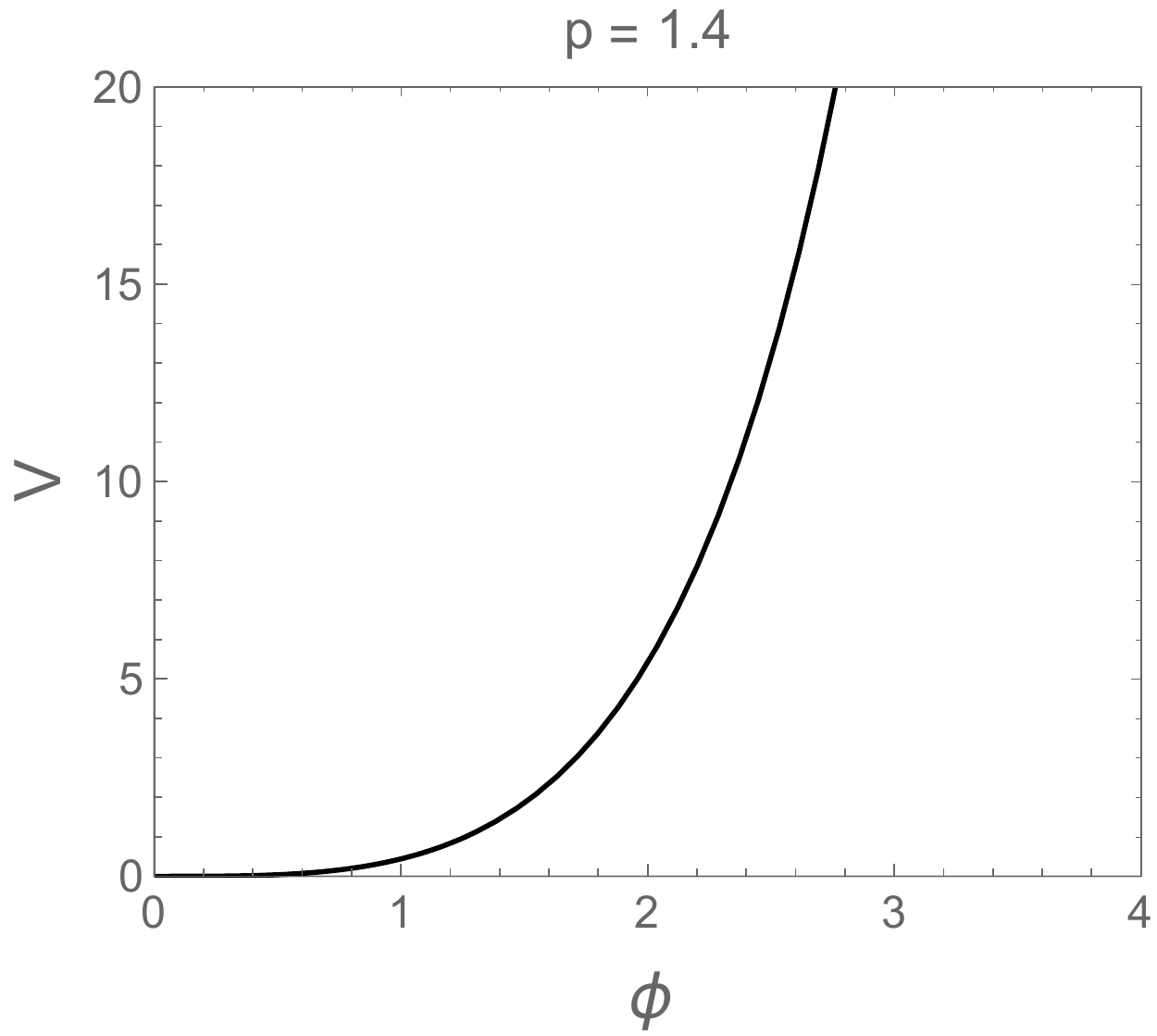}  
\hspace{0.3cm}
\includegraphics[height=4cm]{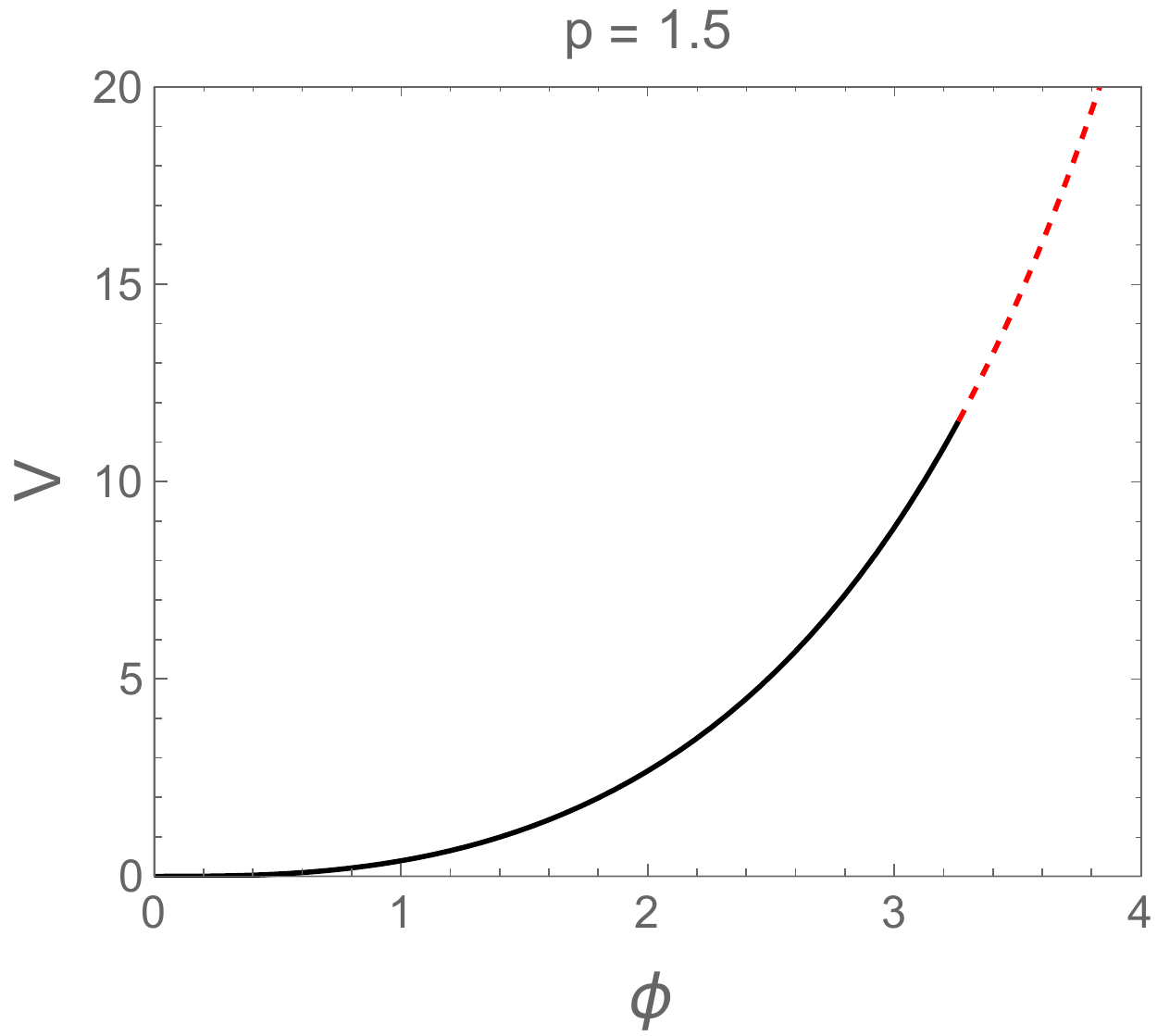}  
\hspace{0.3cm}
\includegraphics[height=4cm]{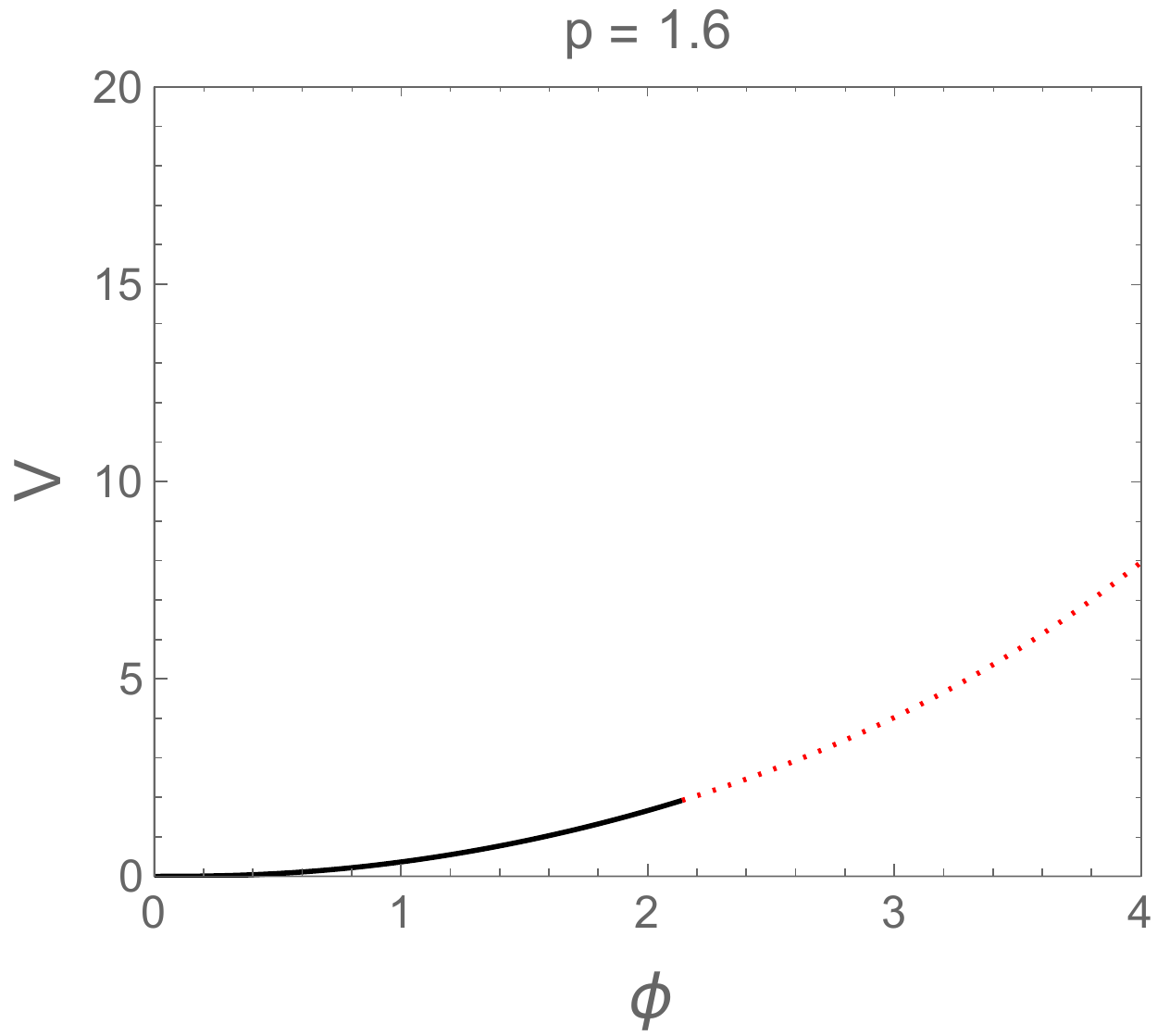}  
\caption{All panels present the Einstein frame potentials for the \eqref{eq:model1} model for $p \in \{1.4,1.5,1.6\}$ (left, middle and right panel respectively).  $c = 1$. Black (red dotted) parts of the potential represent regions, which satisfy (do not satisfy) Eq. (\ref{eq:conditionfRvac}). For $p \lesssim 1.45$ \eqref{eq:conditionfRvac} is satisfied for all values of $R$.}
\label{fig:VfR}
\end{figure}

For most of the realistic $f(R)$ one satisfies $Rf_R < 2f$ (or $Rf_R > 2f$) only for a given range of $R$. Nevertheless one can still find an example of $f(R)$ theory, which satisfy \eqref{eq:conditionfRvac} together with $Rf_R < 2f$ for all values of $R$. Let us consider a simple example like 
\begin{equation}
f(R) = R + c_2 R^p \, , \label{eq:model1}
\end{equation}
This model satisfies \eqref{eq:conditionfRvac} together with $Rf_R < 2f$ for all $R>0$ for 
\begin{equation}
c_2 > 0 \, , \qquad p \in (1,\alpha) <2\, . \label{eq:c2psolution}
\end{equation}
{Examples of scalar potentials for \eqref{eq:model1} with \eqref{eq:c2psolution} are shown in the Fig. \ref{fig:VfR}.} In addition, the model holds $f_{RR} > 0$, so it is free from the tachyonic instability of perturbations. Furthermore, in \cite{DeFelice:2010aj,Artymowski:2014gea} it was shown that such a model may be responsible for generating inflation for $p \in (\frac{1}{2}(1+\sqrt{3}),2)$. Thus, for any $c < \sqrt{2}$ one {may always find $p$, for which (\ref{eq:conditionfRvac}) is satisfied and which} generates inflation for sufficiently large $R$. Nevertheless, such an inflationary model is either equivalent to the power-law inflation (in the case of $p \lesssim 1.9$) and therefore would not be consistent with a current observational data {(which completely excludes power-law inflation - see Fig. 8 in Ref. \cite{Akrami:2018odb})}, or it is close to Starobinsky inflation, and therefore inconsistent with \eqref{eq:conditionspure} \cite{Artymowski:2014gea}.

The (\ref{eq:model1}) model is an interesting example of an $f(R)$ theory, for which one can easily translate the (\ref{eq:deltaphi}) condition into {$\Delta\phi = \sqrt{3/2}\log(f_R(\Delta R))$. Thus, one obtains} the following limitation for $R$
\begin{equation}
\Delta R \lesssim \left( \frac{e^{\sqrt{\frac{2}{3}}}-1}{p\, c_2} \right)^{\frac{1}{p-1}} \, .
\end{equation}
The $c_2$ coefficient has a dimension of mass of the power of $2-2p$, which means, that $c_2$ could be rewritten as $c_2 = s M^{2-2p}$, where $s = \pm 1$ and $M$ is some mass scale, which suppresses the deviations from GR. Then, one finds
\begin{equation}
\Delta R \lesssim M^2 \left( \frac{e^{\sqrt{\frac{2}{3}}}-1}{s\, p} \right)^{\frac{1}{p-1}} \, .
\end{equation}
Then, in the {$p \to 1^+ \ , \ p \to 1^- \ , \ p \to \infty$ limit one finds $\Delta R \to \infty \ , \ \Delta R \to 0 \ , \ \Delta R \to M^2$}. The upper bound on $R$ is an interesting consequence of the swampland conjecture. It shows that in order to have an $f(R)$ theory, which is valid up to the Planck scale, one needs $M \sim M_p$ or $p-1\ll 1$. 

The (\ref{eq:c2psolution}) solution can be generalized into a broader family of functions. Let us note that every $f(R)$, which is a solution 
\begin{equation}
Rf_R = \tilde{\alpha}(R)f \, , \label{eq:fgeneralization}
\end{equation}
where $\tilde{\alpha}(R) < \alpha$ for all $R$, satisfies the (\ref{eq:ineq1}) inequality. The simple example of a solution of Eq. \eqref{eq:fgeneralization} is 
\begin{equation}
f \propto \frac{R^\alpha}{(1+ b R^{2n})^{\frac{\alpha}{2n}}}\, ,
\end{equation}
which automatically satisfies the $Rf_R < 2f$ assumption.

\subsection{Example of $Rf_R>2f$}

Let us assume a polynomial form of $f(R)$, namely
\begin{equation}
f(R) = \sum_{n=1}^{\infty} \left(a_n R^{-n} +  b_n R^{n} \right)\, . \label{eq:fRpolynomial}
\end{equation}
In such a case one can find constraints on $a_n$ and $b_n$ coefficients, which enable to satisfy \eqref{eq:conditionfRvac} together with $Rf_R>2f$. For any realistic $f(R)$ theory the $Rf_R>2f$ condition is satisfied only for a given range of $R$, since one requires an existence of the GR vacuum around $R = 0$. Nevertheless one can still find a highly non-physical example of an $f(R)$ theory, which satisfies $Rf_R>2f$ for all values of $R$. It is possible only for $c < \sqrt{2/3}$ (case (a)) and requires $a_n = 0$ for all $n$ together with
\begin{equation}
b_n = 0 \, \quad \text{for} \quad n < \gamma \, , \quad b_n > 0   \quad \text{for} \quad n > \gamma
\end{equation}
Thus for $c=0.5$ a viable theory is $f(R)=b_4R^4+\cdots$ where the dots indicate higher powers.

\subsection{Range of {theoretical} viability of particular models}

In this subsection, we present several well known $f(R)$ theories. For each one of them, we present the allowed range of $R$, which is consistent with the swampland conjecture. 

\begin{itemize}
\item[i)]Let us start from the Starobinsky model defined by
\begin{equation}
f(R) = R + \frac{R^2}{6M^2} \, .
\end{equation}
Then, the vacuum version of the (\ref{eq:conditionfR}) inequality is satisfied for
\begin{equation}
R \in (-R_c,R_c), \qquad \text{where} \qquad R_c = \frac{2}{c}\sqrt{6}M^2  \, .
\end{equation}
Noted, that the allowed values of $R$ are below the scale of inflation{, which means that Starobinsky inflation is highly inconsistent with the Swampland conjecture}.

\item[ii)] A simple generalization of a Starobinsky model is the 
\begin{equation}
f(R) = R + c_1 R^2 + c_2 R^2\log(R) \, 
\end{equation}
model \cite{Ben-Dayan:2014isa}. For $c_2 < 0$ one can obtain a slope instead of a plateau in $V$. The slope is steep enough to satisfy (\ref{eq:conditionfRvac}) for any value of $R$ for $-c_2 \gtrsim c_1(2-\alpha)$, where $\alpha$ is defined in the Eq. \eqref{eq:Vphipositive}. Note that in such a case $V(\phi)$ would be so steep that it would not support inflation. For some range of $R$ one can also satisfy \eqref{eq:conditionfRvac}  for $c_2>0$. In such a case the potential has a local maximum at $R = \exp(-(c_1+c_2)/c_2)$. At the $V_\phi > 0$ slope (i.e. between the minimum and the maximum) the \eqref{eq:conditionfRvac} may be satisfied, but at the $Rf_R > 2f$ slope it is practically impossible. In order to obtain any region of the $V_\phi < 0$ slope with a desired steepness of the Einstein frame potential, one requires
\begin{equation}
c_1<\frac{c_2}{3 c} \left(\sqrt{6}-3 c+3 c \log \left(\sqrt{\frac{3}{2}} \, c\, c_2\right)\right) \, .
\end{equation}
Unless one assumes $c_2 \sim \mathcal{O}(1)$, the condition above leads to negative values of $c_1$, which are excluded in this theory due to possible quantum instabilities for certain values of $R$.

\item[iii)] The swampland conjecture may also limit models of dark energy. One of the simplest examples of DE $f(R)$ model is \cite{Amendola:2006we} 
\begin{equation}
f(R) = R - c_2 R^n \, ,
\end{equation}
where $c_2 > 0$ and $n \in (0,1)$. In order to avoid $f_R<0$ one requires $R > (c_2 n)^{\frac{1}{1-n}}$. This model satisfies $Rf_R \lessgtr 2f$ only in a particular range of $R$, namely 
\begin{equation}
2f - Rf_R > 0 \qquad \text{for} \qquad R \in (0,(c_2(2-n))^{\frac{1}{1-n}}) \, .
\end{equation}
One can show, that (\ref{eq:ghost}) and (\ref{eq:conditionfR}) are satisfied for any
\begin{equation}
 R > (c_2(2-n+\sqrt{\frac{3}{2}}c \, (1-n)))^{\frac{1}{1-n}} \quad \text{or} \quad R < (c_2(2-n-\sqrt{\frac{3}{2}}c \, (1-n)))^{\frac{1}{1-n}} \, , 
\end{equation}
where we have assumed $c < (2-n)/(1-n)$. The bound on $R$ comes from the fact that $V$ obtains a non-zero, positive value of vacuum evergy, which means that $V_\phi/V \to 0$ around the minimum. Therefore the model cannot play the role of a DE model. {The model presented in here does not take into account the presence of other fields or matter courses in the Universe. As shown in \eqref{eq:Veff} the effective potential may be significantly modified by the presence of e.g. dust.  From the point of view of the late time evolution of the Universe, the scalaron shall always stay in the close vicinity of its de Sitter minimum. The position of the minimum and the steepness of the potential around it strongly depends on $\rho^\star$. Nevertheless such a modification shall not change  the  conclusion. If one waits long enough all sources of matter components will redshift to the point when $V_\phi^{eff}/V^{eff} \sim V_\phi/V$, as in the vacuum case, and $V_\phi^{eff} = 0$ at the minimum. Hence, these $f(R)$ DE models are excluded.} 
\end{itemize}
To summarize this section, it is clear that most $f(R)$ models designated to give rise to an inflationary period or dark energy models are either in discord with the swampland conjecture or are in discord with observational data.

\section{Conditions for $V_{\phi\phi}$} \label{sec:VphiphifR}

Alternatively one can investigate the following conditions in the vacuum case
\begin{equation}
V > 0 \, ,\qquad \frac{V_{\phi\phi}}{V} < -\tilde{c} \, , \label{eq:conditions2}
\end{equation}
where $\tilde{c}>0$ and of order of unity. Fulfilling this condition will allow some maxima of $V(\phi)$ to be consistent with the swampland conjecture. We want to emphasize that the \eqref{eq:conditions2} inequality may only be satisfied locally. A viable $f(R)$ theory must contain a GR minimum, around which $V_{\phi\phi}>0$. 

In the language of the $f(R)$ function and its derivatives the $V_{\phi\phi}$ term is equal to
\begin{equation}
V_{\phi\phi} = \frac{d}{d\phi}V_\phi =  \sqrt{\frac{2}{3}}\frac{f_R}{f_{RR}}\frac{d}{dR}\left( \sqrt{\frac{2}{3}}\frac{f_R}{f_{RR}}V_R\right) = \frac{1}{3f_R^2f_{RR}}(f_R^2+Rf_{RR}f_R-4f_{RR}f) \, .
\end{equation}
The condition (\ref{eq:conditions2}) is therefore equivalent to
\begin{equation}
\frac{1}{f_{RR}}(f_R^2+Rf_{RR}f_R-4f_{RR}f)<-\frac{3}{2}\tilde{c}(Rf_R-f) \, ,
\end{equation}
which gives
\begin{equation}
\left(1+\frac{3}{2}\tilde{c}\right)Rf_{RR}f_R - \left(\frac{3}{2}\tilde{c}+4\right)f_{RR}f + f_R^2 < 0 \, , \label{eq:Vphiphiconstrain}
\end{equation}
First of all we want to look for a form of $f(R)$, for which (\ref{eq:conditions2}) would be satisfied for all $R$. The equation $V_{\phi\phi} = - \tilde{c}\, V $ has an analytical solution of the form of
\begin{equation}
f(R) = c_2 R^p \, , \label{eq:Rpsolution}
\end{equation}
where $p$ is a solution of the following equation
\begin{equation}
 p^2\left(1+\frac{3}{2} \tilde{c}\right) - p \left(3 \tilde{c}+4\right) + \frac{3}{2} \tilde{c} + 4 = 0 \quad \Rightarrow \quad p_{1,2} = \frac{3\tilde{c}+4 \pm i\sqrt{6\tilde{c}}}{2+3\tilde{c}} \, . 
\end{equation}
The solutions are complex and therefore unphysical. In fact it shows that there is no real solution of \eqref{eq:conditions2} of the form of $R^p$. This {may be} obvious, since $R^p$ in the Einstein frame leads to $V \propto \exp(\sqrt{2/3}\phi\frac{2-p}{p-1})$, which for real values of $p$ can never give negative $V_{\phi\phi}/V$. 

The (\ref{eq:conditions2}) condition may be satisfied around maxima of the Einstein frame potential. A typical model in an $f(R)$ framework, which contains an Einstein frame maximum is (\ref{eq:model1}). The Einstein frame potential takes the form of
\begin{equation}
V \propto e^{-2\sqrt{\frac{2}{3}}\phi}\left(e^{\sqrt{\frac{2}{3}}\phi} - 1\right)^\frac{n}{n-1} \, .
\end{equation}
One can show that there exists $\phi$, for which the $V_{\phi\phi} < -\tilde{c} \, V$ condition can be satisfied if
\begin{equation}
n > \frac{2}{9}(3\tilde{c} + 8) \, .
\end{equation}
Since $\tilde{c} \sim \mathcal{O}(1)$ one requires $n$ to be at least slightly bigger than 2. The result is quite intuitive, since for $n > 2$ one finds $V \to e^{-\sqrt{\frac{2}{3}}\frac{n-2}{n-1}\phi}$ for $\phi \to \infty$, which gives a maximum of $V$ between the GR limit (which is $\phi \to 0$) and a run-away vacuum in $\phi \to \infty$. One can find that for any $n$ the inequality \eqref{eq:conditions2} for model \eqref{eq:model1} is satisfied for
\begin{equation}
R_{-} < R <  R_{+}\, , \quad \text{where}\quad R_{\pm}=\left(\left(\frac{n}{2}(3n-5)\pm\frac{n-1}{2}\sqrt{n(9n-6\tilde{c}-16)}\right)c_2\right)^{\frac{-1}{n-1}} \, . \label{eq:Rpm}
\end{equation}
One can see if the $R\in \left(R_{-},R_{+}\right)$ overlaps with regions allowed by the $V_\phi > c\, V$ condition and therefore if in the case of the \eqref{eq:model1}  model one can satisfy \eqref{eq:conditionspure} or \eqref{eq:conditionspure2} for all $R$. The \eqref{eq:conditionfRvac} at the $Rf_R > 2f$ slope is satisfied for
\begin{equation}
R > \left(\frac{\gamma-1}{(n-\gamma)c_2}\right)^{\frac{1}{n-1}} \, , \label{eq:Rminmodel1}
\end{equation}
From Eqs. (\ref{eq:Rpm},\ref{eq:Rminmodel1}) one finds, that the overlap between $V_\phi > c\, V$ and $V_{\phi\phi} < -\tilde{c}\, V$ regions may occur only for $c < \sqrt{2/3}$ and
\begin{equation}
n > \frac{2\gamma^2}{(\gamma-1)(3\tilde{c}+8-(2+3\tilde{c})\gamma)} \, .
\end{equation}
Such a positive $n$ may exist if
\begin{equation}
c < \frac{4-3\tilde{c}}{3\sqrt{6}} \, ,
\end{equation}
which is mathematically possible, but highly unrealistic from the physical point of view. For instance even if $\tilde{c}$ is as small as $1/3$ one requires $c < 1/\sqrt{6}$ in order to provide the overlap of two region, which satisfy \eqref{eq:conditionfRvac} and \eqref{eq:conditions2}. Therefore, for the \eqref{eq:model1} model with $n>2$ one should always expect a region in the Einstein frame potential, which is not steep enough in order to be consistent with the swampland conjecture. 

Examples of \eqref{eq:model1} model, for which the (\ref{eq:conditions2}) conditions are satisfied around a local maximum of $V$ is presented in the right panel of the Fig. \ref{fig:VBD}
\\*

In the non-vacuum case one finds $V_{\phi\phi}^{eff} = V_{\phi\phi} + (1-3w)^2 \rho_m^E/6$. The contribution of matter fields is always positive and therefore the presence of matter is making it harder to satisfy the $V_{\phi\phi}^{eff}/V^{eff} < -\tilde{c}$ condition

\section{The Brans-Dicke generalization of the Swampland conjecture} \label{sec:BD}

The $f(R)$ theory is a particular example of a broader family of theories of modified gravity, called scalar-tensor theory. Its action in the Jordan frame is the following
\begin{equation}
S = \int d^4x \sqrt{-g} \left[\frac{1}{2}R \, F(\varphi) + \frac{K(\varphi)}{2}(\partial \varphi)^2 - U(\varphi)\right] + S_m \, ,
\end{equation}
where $\varphi$ is a Jordan frame field, {$F(\varphi)$ is a non-minimal coupling to gravity}, $K$ determines the kinetic term and $U$ is a Jordan frame potential. After the transformation to the Einstein frame one obtains
\begin{equation}
S = \int d^4x\sqrt{-g^E} \left[ \frac{M_p^2}{2}R^E - \frac{1}{2}(\partial \phi)^2 - V(\phi) \right] + S_m \, ,
\end{equation}
where the Einstein frame field $\phi$ and potential $V$ are defined by
\begin{equation}
\left(\frac{d\phi}{d\varphi}\right)^2 = K^E(\varphi) = \frac{3}{2}\left(\frac{F_\varphi}{F}\right)^2 + \frac{K}{F} \, ,\qquad V = \frac{U}{F^2} \, .
\end{equation}
Thus, the Eq. (\ref{eq:conditionspure}) is equivalent to
\begin{equation}
|FU_\varphi - 2F_\varphi U| > c \,  U F \sqrt{K^E} \, . \label{eq:conditionST}
\end{equation}
In the case of Brans-Dicke theory{, which is a sub-case of the scalar-tensor theory,} one finds
\begin{equation}
F = \varphi \, ,\qquad K = \frac{\omega_\text{\tiny BD}}{\varphi} \quad \Rightarrow \quad K^E = \frac{\beta}{2\varphi^2}  \quad \Rightarrow \quad \phi = \frac{\beta}{2}\log \varphi \, , \label{eq:BD}
\end{equation}
where $\beta = 2\omega_\text{\tiny BD} + 3$. The {$\omega_\text{\tiny BD} = 0$} case is fully equivalent to the $f(R)$ theory. In order to obtain classical and quantum stability one requires $\varphi > 0$ and $U_{\varphi\varphi} > 0$.

In the case of Brans-Dicke theory the condition $\Delta\phi \lesssim 1$ does not depend on the form of the potential. From Eq. (\ref{eq:BD}) one finds
\begin{equation}
\Delta \varphi = e^{\sqrt{\frac{2}{\beta}}} \, ,
\end{equation}
which means, that $\Delta \varphi \to 1$ or $\Delta \varphi \to \infty$ for $\beta \to \infty$ or $\beta \to 0^+$. Like in the case of $f(R)$ theory the limit on $\phi$ corresponds to a very wide range of allowed values of the Jordan frame quantities.

\subsection{The $FU_\varphi > 2F_\varphi U$ case}

From Eqs. (\ref{eq:conditionST},\ref{eq:BD}) one finds
\begin{equation}
\varphi U_\varphi - \left(2+\sqrt{\frac{\beta}{2}}\, c\right)U > 0 \, , \label{eq:BDineq1}
\end{equation}
which gives
\begin{equation}
U(\varphi) > U(a) \left( \frac{\varphi}{a} \right)^{2+\sqrt{\beta/2}\, c} \, .
\end{equation}
Note, that every solution of
\begin{equation}
\varphi U_\varphi - 2U = \sqrt{\frac{\beta}{2}}U\mathcal{C}(\varphi)\, ,  \label{eq:BDgeneral1}
\end{equation}
where $\mathcal{C} > c$ for all $\varphi$, satisfies the inequality (\ref{eq:BDineq1}). The solution of the Eq. (\ref{eq:BDgeneral1}) take the form of
\begin{equation}
U \propto \exp\left(\int\frac{d\varphi}{\varphi}\left(2+\sqrt{\frac{2}{\beta}}\mathcal{C}(\varphi)\right)\right) \, .
\end{equation}
A simple example of such a model would be 
\begin{equation}
\mathcal{C} = c+\alpha \varphi^n \quad \Rightarrow \quad U \propto \varphi^{2+\sqrt{\beta/2}\, c} \exp\left(\frac{\alpha}{n}\sqrt{\frac{\beta}{2}}\varphi^n\right) \, , \label{eq:ctilde}
\end{equation}
where $\alpha > 0$.This solution automatically satisfies the $FU_\varphi > 2F_\varphi U$ assumption.

\subsection{The $FU_\varphi < 2F_\varphi U$ case}

From Eqs. (\ref{eq:conditionST},\ref{eq:BD}) one finds
\begin{equation}
 \left(2-\sqrt{\frac{\beta}{2}}\, c\right)U - \varphi U_\varphi > 0 \, , \label{eq:BDineq2}
\end{equation}
which gives
\begin{equation}
U(\varphi) < U(a) \left( \frac{\varphi}{a} \right)^{2-\sqrt{\beta/2}\, c} \, . \label{eq:BDineq2solgen}
\end{equation}
Following the logic presented in (\ref{eq:BDgeneral1}-\ref{eq:ctilde}) one find, that 
\begin{equation}
U \propto \varphi^{2-\sqrt{\beta/2}\, c} \exp\left(-\frac{\alpha}{n}\sqrt{\frac{\beta}{2}}\varphi^n\right) \, , \label{eq:ctilde2}
\end{equation}
automatically satisfies \eqref{eq:BDineq2}. {Both \eqref{eq:ctilde} and \eqref{eq:ctilde2} models produce the Einstein frame potential, which is steeper than the power-law inflationary scenario. One can see that by noticing that $U \propto \varphi^{2 \pm \sqrt{\beta/2}\, c}$ gives $V \propto e^{\pm c \phi}$. Thus, none of the models presented in this section is actually consistent with the Planck data \cite{Akrami:2018odb}.} We want to note the crucial difference between $f(R)$ theory and its Brans-Dicke generalization. In the BD theory one may easily obtain a solution for $|V_\phi| > c\, V$ with $V_\phi < 0$. We also do not need to distinguish between different values of $c$ in order to find the allowed range of $U(\varphi)$ functions.

\subsection{The non-vacuum case of Brans-Dicke theory}

In the Brans-Dicke theory the effective potential takes the form of 
\begin{equation}
V^{eff}(\phi) = V(\phi) + e^{-\frac{1}{\sqrt{2\beta}}(1-3w)\phi} \rho^\star \, 
\end{equation}
which means that for the domination of matter fields one finds 
\begin{equation}
\frac{|V^{eff}_\phi|}{V^{eff}} \simeq \frac{1}{\sqrt{2\beta}}|1-3\omega| \, .
\end{equation}
For $\beta < \frac{1}{2c^2}(1-3w)^2$ one obtains $|V^{eff}_\phi|/V^{eff} > c$ and therefore the condition for swampland conjecture is satisfied. Note that unlike the $f(R)$ case this can be done for any $w \neq 1/3$.

\section{Condition for $V_{\phi\phi}$ in Brans-Dicke theory} \label{sec:BDVphiphi}

The condition (\ref{eq:conditions2}) can be expressed in terms of scalar-tensor theory as
\begin{equation}
\frac{V_{\phi\phi}}{V} = \frac{1}{UK^EF^2}\left(F^2 U_{\varphi\varphi}-4F_\varphi F U_\varphi - 2 F_{\varphi\varphi}FU+6F_\varphi^2U - F\frac{K^E_\varphi}{2K^E}(F U_\varphi-2F_\varphi U)\right) < - \tilde{c} \, ,
\end{equation}
where
\begin{equation}
\frac{K^E_\varphi}{2K^E} = \frac{F_{\varphi\varphi}F - F_\varphi^2+F K_\varphi - F_\varphi K}{3F_\varphi^2 + 2FK}
\end{equation}
In the case of Brans-Dicke theory one finds
\begin{equation}
\frac{V_{\phi\phi}}{V} = \frac{1}{\beta U}\left(\varphi^2 U_{\varphi\varphi} - 3 \varphi U_\varphi + 4U\right) \, ,
\end{equation}
which gives the condition
\begin{equation}
\varphi^2 U_{\varphi\varphi} - 3 \varphi U_\varphi + 4U + \tilde{c} \beta U < 0 \, . \label{eq:BDVbisineq}
\end{equation}
As in the $f(R)$ the inequality (\ref{eq:BDVbisineq}) does not have a simple solution. If (\ref{eq:BDVbisineq}) would be an equation, it would have a complex solution of the form of
\begin{equation}
U \propto \varphi^n \, \qquad n = 2 \pm i \sqrt{\tilde{c}\beta} \, . \label{eq:Vphiphisolution}
\end{equation}
For $U \propto \varphi^n$ and positive $U$ the LHS of (\ref{eq:BDVbisineq}) is always positive and therefore such a solution cannot satisfy the $V_{\phi\phi} < - \tilde{c}\, V$ condition. Again, one should expect that, since the $V>0$ condition enforce the existence of minima in the Einstein frame. Unlike in the case of the $f(R)$ theory, the \eqref{eq:conditions2} condition takes the form of the linear differential inequality. In such a case on can construct a real solution of \eqref{eq:BDVbisineq} as a linear combination of solutions from \eqref{eq:Vphiphisolution}, which in the most general case takes the form of
\begin{equation}
U \propto \varphi^2\left(A \sin(\sqrt{\tilde{c}\, \beta}\log\varphi) + B \cos(\sqrt{\tilde{c}\, \beta}\log\varphi)\right) \, ,
\end{equation}
where $A,B$ are constants of integration. Such a model satisfies \eqref{eq:conditions2} for all values of $\varphi$. Unfortunatelly it often leads to $U_{\varphi\varphi} < 0$ and in consequence to tachyonic scalar perturbations.
\\*

\begin{figure}[h!]
\centering
\includegraphics[height=4cm]{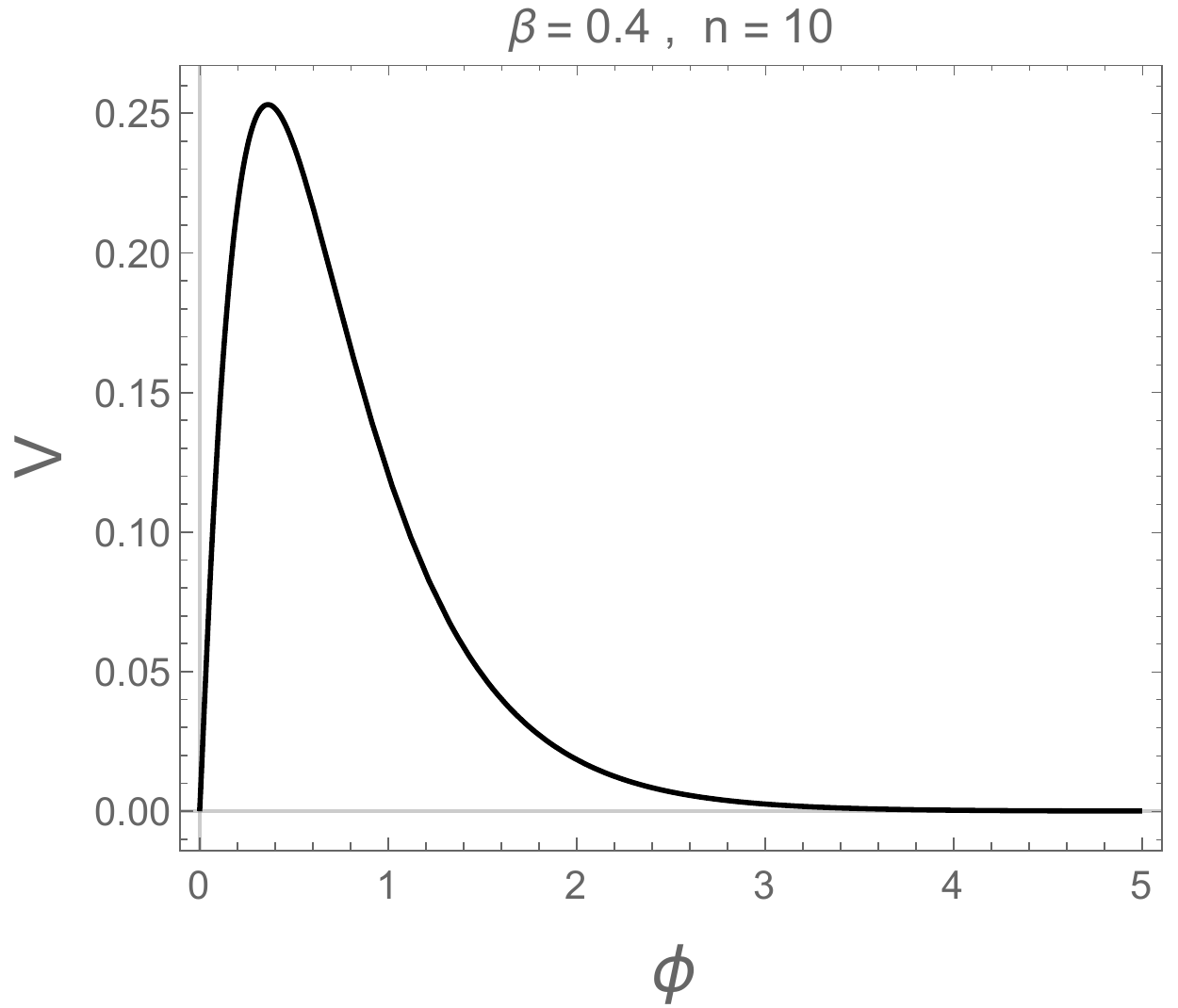}  
\hspace{0.3cm}
\includegraphics[height=4cm]{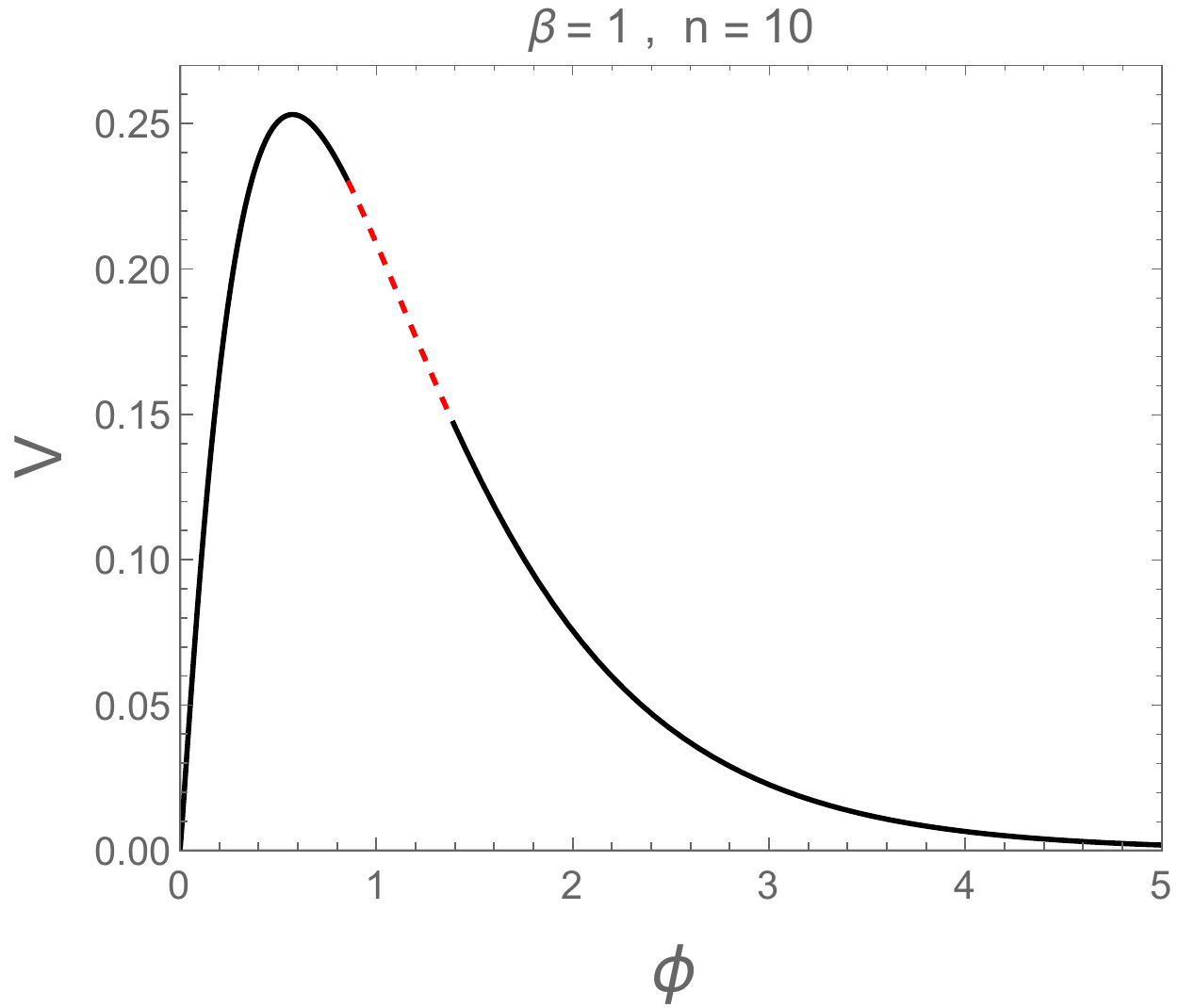}  
\hspace{0.3cm}
\includegraphics[height=4cm]{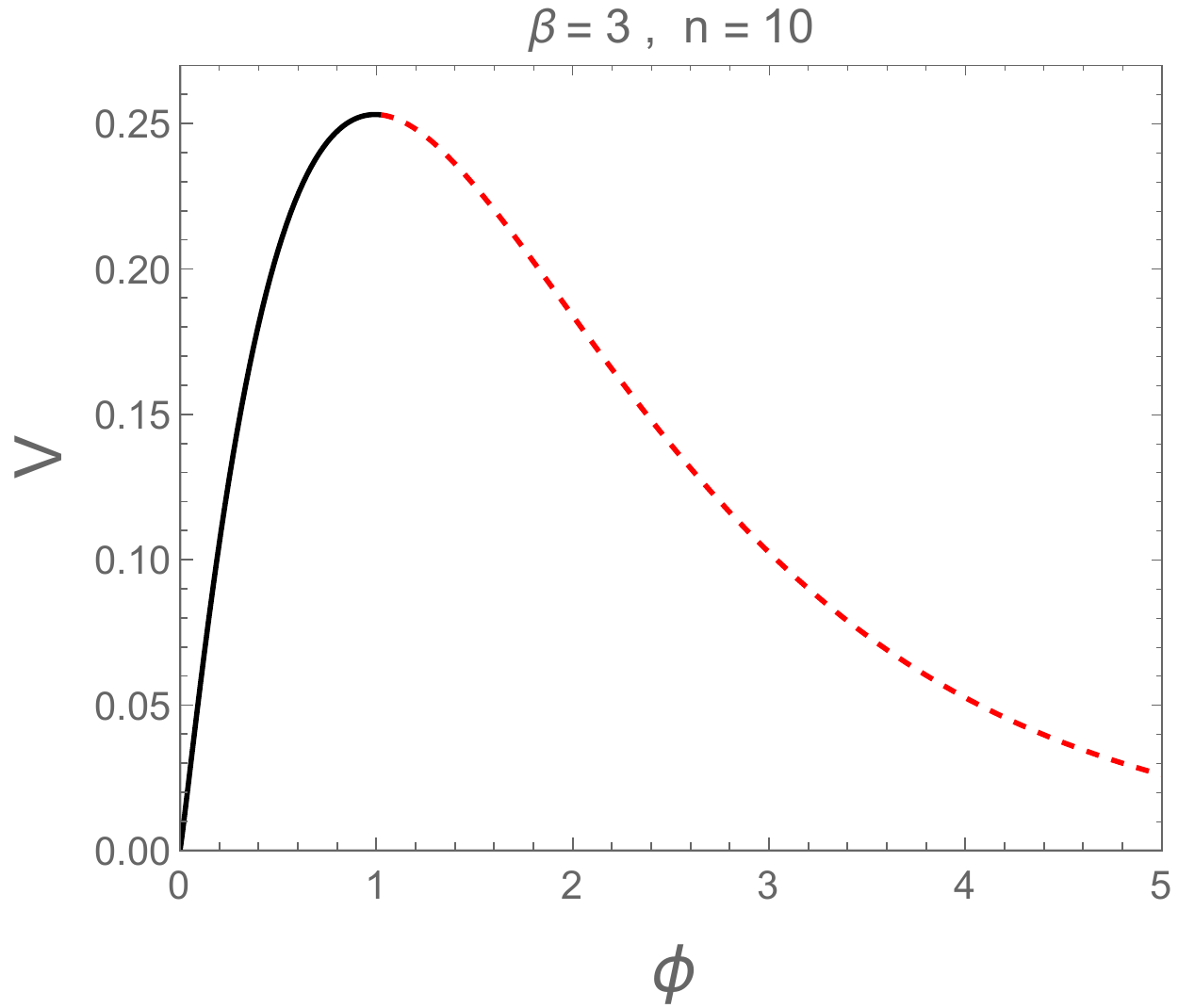}  
\caption{All panels present the Einstein frame potentials for the Brans-Dicke generalization of the $f(R) = R + c_2 R^n$ model for $n=10$ and $\beta \in \{0.4, 1, 3\}$ (left, middle and right panel respectively). The $\beta = 3$ case corresponds to the $f(R)$ theory. We have assumed $c = \tilde{c} = 1$. Black (red) parts of the potential represent regions, which satisfy (do not satisfy) Eq. (\ref{eq:conditionspure}) or (\ref{eq:conditions2}). The (\ref{eq:conditionspure}) condition is satisfied at the steep slopes of potentials, while the (\ref{eq:conditions2}) are satisfied around the local maximum of the potential. The viability of theories presented here is strongly sensitive to values of $c$ and $\tilde{c}$.}
\label{fig:VBD}
\end{figure}

As in the case of $f(R)$ theory, one may obtain some range of the value of the field, for which the $V_{\phi\phi}< - \tilde{c} \, V$ condition is satisfied. In the Fig \ref{fig:VBD} we present an example of a Brans-Dicke generalization of an $f(R)$ theory, for which one obtain regions of the potential, which satisfy either (\ref{eq:conditionspure}) or (\ref{eq:conditions2}).

\section{Conclusions} \label{sec:concl}

In this article we have investigated the swampland conjecture constrains on $f(R)$ and Brans-Dicke theories of modified gravity. We have translated the swampland conjecture into the language of these theories. We have constrained the $f(R)$ theory with respect to the swampland conjecture, the existence of the GR vacuum and the classical and quantum stability of the theory ($f_R > 0$ and $f_{RR} > 0$). We have shown that in the presence of perfect fluid with dominant energy density one may easily satisfy $V_\phi > c\, V$, even if the perfect fluid mimics inflaton or dark energy. Thus, the $f(R)$ scalaron may play a role of subdominant scalar field with a steep potential, which enables consistency between the swampland conjecture and inflation. 

In $FR<2f$ and $FR>2f$ regimes (which correspond to $V_\phi > 0$ and $V_\phi < 0$ respectively) we have shown that in order to satisfy $V_\phi > c \, V$ one requires: (a) $f(R_\star)(R/R_\star) < f(R) < f(R_\star)(R/R_\star)^\alpha$, or (b) $f(R) > f(R_\star)(R/R_\star)^\gamma$, with $\alpha$ and $\gamma$ written in the text. The precise value of $c$ strongly determines the existence of any {theoretical} viable $f(R)$ theory with $Rf_R > 2f$, which is significantly different than in the case of other scalar theories.

Furthermore, we have analyzed few examples of realistic $f(R)$ theories of inflation and dark energy in the context of their consistency with the swampland conjecture. Not surprisingly, the swampland conjecture is strongly inconsistent with $f(R)$ inflation. Specifically we have focused on the $f(R) = R + c_2 R^p$ model. We have proven that only for $p \in (1,\alpha)$ one can satisfy the swampland conjecture for all values of $R$. 

{For DE and the late-time Universe, the exact value of $c$ is crucial for the phenomenological consistency of particular models. We have investigated a popular example of a DE $f(R)$ model, namely $f(R) = R - c_2\, R^p$ with $0<p<1$ and we have concluded that for the late time evolution the scalaron one always ends up in the DS vacuum, which is inconsistent with the swampland conjecture.}

Brans-Dicke {theories} have an additional free parameter comparing to $f(R)$ and therefore it is easier for them to satisfy the swampland conjecture. In the non-vacuum case the domination of matter fields with $\beta < \frac{1}{2c^2} (1-3w)^2$ leads to $|V^{eff}_\phi|/V > c$. Thus, there always exist BD theories with small enough $\beta$, such that for a given equation of state parameter $w$, the swampland conjecture is {satisfied}. In addition, we placed bounds from the swampland conjecture on the vacuum BD theory.

{As in $f(R)$ theory, the second conjecture may only be satisfied locally {(i.e. in a vicinity of some $R$)} for viable theories. Hence, at least in the case of $f(R)$ and BD gravity, the second swampland criterion of $V_{\phi \phi}<-\tilde{c} \, V$ can be valid only in a limited patch of field space, while in most of field space, the first criterion $|V_{\phi}|>c\, V$ has to be fulfilled. As such, both criteria magically conspire to allow for maxima in $f(R)$ or BD gravity, while not  allowing viable inflation or DE domination. 
Thus, the conjecture, while intriguing, is in tension with Inflation and DE. It would have gathered better evidence and power, had it pointed to a specific class or theories that are consistent and phenomenologically viable. Given that almost any quantum gravity theory is expected to have some higher powers of the Ricci scalar, our study shows that such correction will be limited severely by the swampland conjecture, while still not giving rise to inflation or DE. Without further evidence of the refined conjecture, such a coincidence of allowing $f(R)$/BD but disallowing Inflation/DE seems artificial from $f(R)$ or BD gravity point of view, weakening the conjecture's validity. }

\section*{Acknowledgements}

We thank Alexander Vikman for discussion and useful comments.


\begin{thebibliography}{99}

%\cite{DeFelice:2010aj}
\bibitem{DeFelice:2010aj}
  A.~De Felice and S.~Tsujikawa,
  %``f(R) theories,''
  Living Rev.\ Rel.\  {\bf 13} (2010) 3
  %doi:10.12942/lrr-2010-3
  [arXiv:1002.4928 [gr-qc]].
  %%CITATION = doi:10.12942/lrr-2010-3;%%
  %1657 citations counted in INSPIRE as of 12 Nov 2018
  
  %\cite{Sotiriou:2008rp}
\bibitem{Sotiriou:2008rp}
  T.~P.~Sotiriou and V.~Faraoni,
  %``f(R) Theories Of Gravity,''
  Rev.\ Mod.\ Phys.\  {\bf 82} (2010) 451
  doi:10.1103/RevModPhys.82.451
  [arXiv:0805.1726 [gr-qc]].
  %%CITATION = doi:10.1103/RevModPhys.82.451;%%
  %2234 citations counted in INSPIRE as of 28 Jan 2019

%\cite{Starobinsky:1980te}
\bibitem{Starobinsky:1980te}
  A.~A.~Starobinsky,
  %``A New Type of Isotropic Cosmological Models Without Singularity,''
  Phys.\ Lett.\ B {\bf 91} (1980) 99
   [Phys.\ Lett.\  {\bf 91B} (1980) 99]
   [Adv.\ Ser.\ Astrophys.\ Cosmol.\  {\bf 3} (1987) 130].
  %doi:10.1016/0370-2693(80)90670-X
  %%CITATION = doi:10.1016/0370-2693(80)90670-X;%%
  %3892 citations counted in INSPIRE as of 27 Dec 2018
  
  %\cite{Kehagias:2013mya}
\bibitem{Kehagias:2013mya}
  A.~Kehagias, A.~Moradinezhad Dizgah and A.~Riotto,
  %``Remarks on the Starobinsky model of inflation and its descendants,''
  Phys.\ Rev.\ D {\bf 89} (2014) no.4,  043527
  doi:10.1103/PhysRevD.89.043527
  [arXiv:1312.1155 [hep-th]].
  %%CITATION = doi:10.1103/PhysRevD.89.043527;%%
  %107 citations counted in INSPIRE as of 28 Jan 2019
  
    %\cite{Ben-Dayan:2014isa}
\bibitem{Ben-Dayan:2014isa}
  I.~Ben-Dayan, S.~Jing, M.~Torabian, A.~Westphal and L.~Zarate,
  %``$R^2\log R$ quantum corrections and the inflationary observables,''
  JCAP {\bf 1409} (2014) 005
  doi:10.1088/1475-7516/2014/09/005
  [arXiv:1404.7349 [hep-th]].
  %%CITATION = doi:10.1088/1475-7516/2014/09/005;%%
  %25 citations counted in INSPIRE as of 01 Jan 
  
  %\cite{Bamba:2014mua}
\bibitem{Bamba:2014mua}
  K.~Bamba, G.~Cognola, S.~D.~Odintsov and S.~Zerbini,
  %``One-loop modified gravity in a de Sitter universe, quantum-corrected inflation, and its confrontation with the Planck result,''
  Phys.\ Rev.\ D {\bf 90} (2014) no.2,  023525
  doi:10.1103/PhysRevD.90.023525
  [arXiv:1404.4311 [gr-qc]].
  %%CITATION = doi:10.1103/PhysRevD.90.023525;%%
  %34 citations counted in INSPIRE as of 16 Feb 2019
  
  %\cite{Artymowski:2015mva}
\bibitem{Artymowski:2015mva}
  M.~Artymowski, Z.~Lalak and M.~Lewicki,
  %``Inflationary scenarios in Starobinsky model with higher order corrections,''
  JCAP {\bf 1506} (2015) 032
  doi:10.1088/1475-7516/2015/06/032
  [arXiv:1502.01371 [hep-th]].
  %%CITATION = doi:10.1088/1475-7516/2015/06/032;%%
  %17 citations counted in INSPIRE as of 28 Jan 

  %\cite{Artymowski:2014gea}
\bibitem{Artymowski:2014gea}
  M.~Artymowski and Z.~Lalak,
  %``Inflation and dark energy from f(R) gravity,''
  JCAP {\bf 1409} (2014) 036
  doi:10.1088/1475-7516/2014/09/036
  [arXiv:1405.7818 [hep-th]].
  %%CITATION = doi:10.1088/1475-7516/2014/09/036;%%
  %28 citations counted in INSPIRE as of 12 Nov 2018
  
  %\cite{Codello:2014sua}
\bibitem{Codello:2014sua}
  A.~Codello, J.~Joergensen, F.~Sannino and O.~Svendsen,
  %``Marginally Deformed Starobinsky Gravity,''
  JHEP {\bf 1502} (2015) 050
  doi:10.1007/JHEP02(2015)050
  [arXiv:1404.3558 [hep-ph]].
  %%CITATION = doi:10.1007/JHEP02(2015)050;%%
  %37 citations counted in INSPIRE as of 29 Jan 2019
  
  %\cite{Sebastiani:2013eqa}
\bibitem{Sebastiani:2013eqa}
  L.~Sebastiani, G.~Cognola, R.~Myrzakulov, S.~D.~Odintsov and S.~Zerbini,
  %``Nearly Starobinsky inflation from modified gravity,''
  Phys.\ Rev.\ D {\bf 89} (2014) no.2,  023518
  doi:10.1103/PhysRevD.89.023518
  [arXiv:1311.0744 [gr-qc]].
  %%CITATION = doi:10.1103/PhysRevD.89.023518;%%
  %86 citations counted in INSPIRE as of 05 Feb 2019
  
  %\cite{Broy:2016mdc}
\bibitem{Broy:2016mdc} 
  B.~J.~Broy,
  %``Inflation and Effective Shift Symmetries,''
  doi:10.3204/PUBDB-2016-03859
  %%CITATION = doi:10.3204/PUBDB-2016-03859;%%
  %1 citations counted in INSPIRE as of 05 Feb 2019

%\cite{Broy:2014xwa}
\bibitem{Broy:2014xwa} 
  B.~J.~Broy, F.~G.~Pedro and A.~Westphal,
  %``Disentangling the $f(R)$ - Duality,''
  JCAP {\bf 1503}, no. 03, 029 (2015)
  doi:10.1088/1475-7516/2015/03/029
  [arXiv:1411.6010 [hep-th]].
  %%CITATION = doi:10.1088/1475-7516/2015/03/029;%%
  %23 citations counted in INSPIRE as of 05 Feb 2019

  
  %\cite{Akrami:2018odb}
\bibitem{Akrami:2018odb}
  Y.~Akrami {\it et al.} [Planck Collaboration],
  %``Planck 2018 results. X. Constraints on inflation,''
  arXiv:1807.06211 [astro-ph.CO].
  %%CITATION = ARXIV:1807.06211;%%
  %178 citations counted in INSPIRE as of 27 Dec 2018

%\cite{Array:2015xqh}
\bibitem{Array:2015xqh}
  P.~A.~R.~Ade {\it et al.} [BICEP2 and Keck Array Collaborations],
  %``Improved Constraints on Cosmology and Foregrounds from BICEP2 and Keck Array Cosmic Microwave Background Data with Inclusion of 95 GHz Band,''
  Phys.\ Rev.\ Lett.\  {\bf 116} (2016) 031302
  doi:10.1103/PhysRevLett.116.031302
  [arXiv:1510.09217 [astro-ph.CO]].
  %%CITATION = doi:10.1103/PhysRevLett.116.031302;%%
  %472 citations counted in INSPIRE as of 27 Dec 2018
  
  %\cite{Obied:2018sgi}
\bibitem{Obied:2018sgi} 
  G.~Obied, H.~Ooguri, L.~Spodyneiko and C.~Vafa,
  %``De Sitter Space and the Swampland,''
  arXiv:1806.08362 [hep-th].
  %%CITATION = ARXIV:1806.08362;%%
  %20 citations counted in INSPIRE as of 05 Aug 2018
  
  %\cite{Ooguri:2018wrx}
\bibitem{Ooguri:2018wrx}
  H.~Ooguri, E.~Palti, G.~Shiu and C.~Vafa,
  %``Distance and de Sitter Conjectures on the Swampland,''
  Phys.\ Lett.\ B {\bf 788} (2019) 180
  doi:10.1016/j.physletb.2018.11.018
  [arXiv:1810.05506 [hep-th]].
  %%CITATION = doi:10.1016/j.physletb.2018.11.018;%%
  %69 citations counted in INSPIRE as of 21 Feb 2019
  
  %\cite{Agrawal:2018own}
\bibitem{Agrawal:2018own}
  P.~Agrawal, G.~Obied, P.~J.~Steinhardt and C.~Vafa,
  %``On the Cosmological Implications of the String Swampland,''
  Phys.\ Lett.\ B {\bf 784} (2018) 271
  doi:10.1016/j.physletb.2018.07.040
  [arXiv:1806.09718 [hep-th]].
  %%CITATION = doi:10.1016/j.physletb.2018.07.040;%%
  %103 citations counted in INSPIRE as of 28 Jan 2019
  
  %\cite{Garg:2018reu}
\bibitem{Garg:2018reu}
  S.~K.~Garg and C.~Krishnan,
  %``Bounds on Slow Roll and the de Sitter Swampland,''
  arXiv:1807.05193 [hep-th].
  %%CITATION = ARXIV:1807.05193;%%
  %62 citations counted in INSPIRE as of 28 Jan 2019

%\cite{Denef:2018etk}
\bibitem{Denef:2018etk}
  F.~Denef, A.~Hebecker and T.~Wrase,
  %``de Sitter swampland conjecture and the Higgs potential,''
  Phys.\ Rev.\ D {\bf 98} (2018) no.8,  086004
  doi:10.1103/PhysRevD.98.086004
  [arXiv:1807.06581 [hep-th]].
  %%CITATION = doi:10.1103/PhysRevD.98.086004;%%
  %56 citations counted in INSPIRE as of 28 Jan 2019
  
  %\cite{Roupec:2018mbn}
\bibitem{Roupec:2018mbn}
  C.~Roupec and T.~Wrase,
  %``de Sitter extrema and the swampland,''
  Fortsch.\ Phys.\  {\bf 2018} 1800082
  doi:10.1002/prop.201800082
  [arXiv:1807.09538 [hep-th]].
  %%CITATION = doi:10.1002/prop.201800082;%%
  %43 citations counted in INSPIRE as of 29 Jan 2019
  
  %\cite{Motaharfar:2018zyb}
\bibitem{Motaharfar:2018zyb}
  M.~Motaharfar, V.~Kamali and R.~O.~Ramos,
  %``Warm way out of the Swampland,''
  arXiv:1810.02816 [astro-ph.CO].
  %%CITATION = ARXIV:1810.02816;%%
  %22 citations counted in INSPIRE as of 29 Jan 2019
  
  %\cite{Ben-Dayan:2018mhe}
\bibitem{Ben-Dayan:2018mhe}
  I.~Ben-Dayan,
  %``Draining the Swampland,''
  arXiv:1808.01615 [hep-th].
  %%CITATION = ARXIV:1808.01615;%%
  %28 citations counted in INSPIRE as of 29 Jan 2019
  
  %\cite{Andriot:2018wzk}
\bibitem{Andriot:2018wzk}
  D.~Andriot,
  %``On the de Sitter swampland criterion,''
  Phys.\ Lett.\ B {\bf 785} (2018) 570
  doi:10.1016/j.physletb.2018.09.022
  [arXiv:1806.10999 [hep-th]].
  %%CITATION = doi:10.1016/j.physletb.2018.09.022;%%
  %62 citations counted in INSPIRE as of 01 Feb 2019
  
  %\cite{Ooguri:2016pdq}
\bibitem{Ooguri:2016pdq}
  H.~Ooguri and C.~Vafa,
  %``Non-supersymmetric AdS and the Swampland,''
  Adv.\ Theor.\ Math.\ Phys.\  {\bf 21} (2017) 1787
  doi:10.4310/ATMP.2017.v21.n7.a8
  [arXiv:1610.01533 [hep-th]].
  %%CITATION = doi:10.4310/ATMP.2017.v21.n7.a8;%%
  %100 citations counted in INSPIRE as of 29 Jan 
  
  %\cite{Kehagias:2018uem}
\bibitem{Kehagias:2018uem}
  A.~Kehagias and A.~Riotto,
  %``A note on Inflation and the Swampland,''
  Fortsch.\ Phys.\  {\bf 66} (2018) 100052
  doi:10.1002/prop.201800052
  [arXiv:1807.05445 [hep-th]].
  %%CITATION = doi:10.1002/prop.201800052;%%
  %51 citations counted in INSPIRE as of 29 Jan 2019
  
  %\cite{Raveri:2018ddi}
\bibitem{Raveri:2018ddi}
  M.~Raveri, W.~Hu and S.~Sethi,
  %``Swampland Conjectures and Late-Time Cosmology,''
  arXiv:1812.10448 [hep-th].
  %%CITATION = ARXIV:1812.10448;%%
  %4 citations counted in INSPIRE as of 20 Feb 2019

%\cite{Akrami:2018ylq}
\bibitem{Akrami:2018ylq}
  Y.~Akrami, R.~Kallosh, A.~Linde and V.~Vardanyan,
  %``The landscape, the swampland and the era of precision cosmology,''
  Fortsch.\ Phys.\  {\bf 2018} 1800075
  doi:10.1002/prop.201800075
  [arXiv:1808.09440 [hep-th]].
  %%CITATION = doi:10.1002/prop.201800075;%%
  %60 citations counted in INSPIRE as of 20 Feb 2019
  
  %\cite{Kinney:2018nny}
\bibitem{Kinney:2018nny}
  W.~H.~Kinney, S.~Vagnozzi and L.~Visinelli,
  %``The Zoo Plot Meets the Swampland: Mutual (In)Consistency of Single-Field Inflation, String Conjectures, and Cosmological Data,''
  arXiv:1808.06424 [astro-ph.CO].
  %%CITATION = ARXIV:1808.06424;%%
  %45 citations counted in INSPIRE as of 20 Feb 2019

%\cite{Grimm:2018ohb}
\bibitem{Grimm:2018ohb} 
  T.~W.~Grimm, E.~Palti and I.~Valenzuela,
  %``Infinite Distances in Field Space and Massless Towers of States,''
  arXiv:1802.08264 [hep-th].
  %%CITATION = ARXIV:1802.08264;%%
  %16 citations counted in INSPIRE as of 05 Aug 2018
  
    %\cite{Achucarro:2018vey}
\bibitem{Achucarro:2018vey}
  A.~Ach{\'u}carro and G.~A.~Palma,
  %``The string swampland constraints require multi-field inflation,''
  arXiv:1807.04390 [hep-th].
  %%CITATION = ARXIV:1807.04390;%%
  %56 citations counted in INSPIRE as of 01 Feb 2019
  
  %\cite{Andriot:2018mav}
\bibitem{Andriot:2018mav}
  D.~Andriot and C.~Roupec,
  %``Further refining the de Sitter swampland conjecture,''
  Fortsch.\ Phys.\  {\bf 67} (2019) no.1-2,  1800105
  doi:10.1002/prop.201800105
  [arXiv:1811.08889 [hep-th]].
  %%CITATION = doi:10.1002/prop.201800105;%%
  %11 citations counted in INSPIRE as of 17 Feb 2019
  
  %\cite{Amendola:2006we}
\bibitem{Amendola:2006we}
  L.~Amendola, R.~Gannouji, D.~Polarski and S.~Tsujikawa,
  %``Conditions for the cosmological viability of f(R) dark energy models,''
  Phys.\ Rev.\ D {\bf 75} (2007) 083504
  doi:10.1103/PhysRevD.75.083504
  [gr-qc/0612180].
  %%CITATION = doi:10.1103/PhysRevD.75.083504;%%
  %523 citations counted in INSPIRE as of 20 Nov 2018



\end{thebibliography}
\end{document}